\documentclass[a4paper,11pt]{article}
\pdfoutput=1 

\usepackage{jcappub} 

\usepackage[utf8]{inputenc}

\usepackage{graphicx}
\usepackage{dcolumn}
\usepackage{bm}

\def\be{\begin{equation}}
\def\te{\end{equation}}
\def\ee{\end{equation}}
\def\ba{\begin{eqnarray}}
\def\bea{\begin{eqnarray}}
\def\nn{\nonumber\\}
\def\tea{\end{eqnarray}}
\def\ea{\end{eqnarray}}
\def\eea{\end{eqnarray}}

\usepackage{amsmath}
\usepackage{amssymb}

\usepackage{hyperref} 

\title{\boldmath Relativistic viscous effects on the primordial gravitational waves spectrum}

\author{Nahuel Mirón-Granese}

\affiliation{Departamento de Física \\ Facultad de Ciencias Exactas y Naturales \\ Universidad de Buenos Aires \\ Ciudad de Buenos Aires, Argentina}

\emailAdd{nahuelmg@df.uba.ar}

\abstract{We study the impact of the viscous effects of the primordial plasma on the evolution of the primordial gravitational waves (pGW) spectrum from Inflation until today, considering a self-consistent interaction that incorporates the back-reaction of the GW into the plasma. We use a relativistic causal hydrodynamic framework with a positive entropy production based on a Second-Order Theory (SOT) in which the viscous properties of the fluid are effectively described by a new set of independent variables. We study how the spin-2 modes typical of SOTs capture the simplest GW-fluid viscous interaction to first order. We consider that all non-ideal properties of the primordial plasma are due to an extra effectively massless self-interacting scalar field whose state becomes a many-particles one after Reheating and for which an effective fluid description is suitable. We numerically solve the evolution equations and explicitly compute the current GW spectrum obtaining two contributions. On the one hand we have the viscous evolution of the pGW: For the collision-dominated regime the GW source becomes negligible while in the collisionless limit there exists an absorption of the pGW energy due to the damping effect produced by the free-streaming spin-2 modes of the fluid and driven by the expansion of the Universe. The latter effect is characterized by a relative amplitude decrease of about 1 to 10 \% with respect to the GW free evolution spectrum. On the other hand we get the GW production due to the decay of the initial spin-2 fluctuations of the fluid that is negligible compared with the above-mentioned contribution.

This SOT framework captures the same qualitative effects on the evolution of GW coupled to matter reported in previous works in which a kinetic theory approach has been used.}

\begin{document}
\maketitle
\flushbottom

\section{\label{sec:intro}Introduction}

The standard model of cosmology assumes that the Universe undergoes an inflationary phase of exponential expansion in its earliest stages. During this period of rapid expansion, quantum fluctuations of the metric and matter fields were frozen into super-Hubble classical perturbations which in turn provide the initial conditions for the Universe we observe today. In this work we are interested in the tensor (spin-2) metric perturbations, the so-called primordial gravitational waves (pGW). We will analyze a non-trivial effect on its evolution due to the presence of a viscous primordial plasma.

Accurate knowledge of the current spectrum of the pGW could be a powerful tool to study the very early Universe and high-energy physics when contrasting with an eventual detection by, for example, Pulsar Timing Arrays or Gravitational Waves Interferometers (see \cite{reviewcaprinifigueroa2018,maggiore2018,giovannini2020} and references therein). Even more, an indirect observation might be achieved through the primordial B modes of the CMB polarization that many experiments, as QUBIC \cite{qubic}, are intended to detect \cite{reviewcaprinifigueroa2018,maggiore2018}. It is well-known that the spectrum of pGW at the time of recombination determines the existence of these particular polarization modes. Any observation related to the pGW would provide a fundamental evidence for the inflationary model and would allow us to shed light on the properties of the very early Universe.

The evolution of the pGW is dictated by the transverse and traceless (TT) projection of the linearized Einstein's equations and it depends on both the scale factor dynamics $a(t)$ and the TT projection of the anisotropic stress tensor $\Pi^{\rm TT}_{ij}$ of matter, if present, acting as a source. The standard pGW spectrum has been studied in \cite{watanabe2006,kuroyanagi2009,saikawa2018} where it was shown that even under free-evolution, non trivial effects on the spectrum arise due to changes in the scale factor $a(t)$ through the different eras (radiation, matter and dark energy) or the decay of relativistic degrees of freedom. The production and evolution of GW due to non-standard equations of state of the Universe \cite{relativisticfreestreamingases2016,boyle2008} and due to different inflationary and post-inflationary scenarios \cite{sa2008,sa2010} have also been studied.

Among the effects regarding the non-free evolution of pGW, i.e. $\Pi^{\rm TT}_{ij}\neq0$, we mention the damping due to free-streaming particles \cite{weinberg2004,watanabe2006,matarrese2008,saikawa2018,boyle2008,kuroyanagi2009,repko2013,relativisticfreestreamingases2016,hook2020} and the absorption of GW while propagating in a viscous medium \cite{hawking1966,weinbergpaper1971,weinberglibro1972,anile1978,madore1973,esposito1971,prasanna1999,dent2013,baym2017,brevik2019,constraints_goswami2017,gravitonsandcdmarxivlu2018,zarei2021}. We also remark the studies about the angular anisotropies of the GW background \cite{contaldi2017, matarrese2020}. On the other hand, GW could also be produced within the cosmological context by other phenomena than Inflation, in which the spin-2 anisotropies $\Pi^{\rm TT}_{ij}$ are entirely responsible for producing the GW, such as that the evolution of scalar fields during (p)reheating \cite{dufaux2007,hyde2015,figueroa2016,lozanovamin2019}, cosmological phase transitions \cite{hindmarsh2014,leitao2016,hindmarsh2017,hindmarsh2020}, second order evolution of scalar perturbations \cite{matarrese1998, noh2004,carbone2005, nakamura2007,wands2007,pertdensbaumann2007,domenech2020,domcke2020} and fluctuating thermal plasmas \cite{delcampoford1988,ghiglieri2015,ghiglieri2020,mcdonough2020}. These GW might also evolve non-trivially once created. For an updated review about cosmological GW and its sources see \cite{reviewcaprinifigueroa2018,maggiore2018,giovannini2020} and references therein.

In the present manuscript we focus our attention on the evolution of the pGW coupled to the primordial viscous plasma through a self-consistent interaction which incorporates the back-reaction of the GWs on the plasma. We use a thermodynamical and causally consistent hydrodynamic framework to study some of the simplest non-ideal effects that might be present in the primordial plasma. We will use the basic ideas developed in \cite{nahuelesteban2018} to set the coupled self-consistent dynamics between the GW and the primordial fluid.

We assume that the primordial plasma can be effectively described as a viscous fluid displaying relativistic and dissipative effects in a consistent way according to the second law of thermodynamics. There exist at least two major schemes in order to describe relativistic and dissipative fluid dynamics. On the one hand we have the First-Order Theories (FOTs) constructed from a straightforward covariantization of the well-known non-relativistic Navier-Stokes equation while keeping unchanged the number of degrees of freedom \cite{romatschkeandromatschke,rezzollazanotti}. On the other hand we have the so-called Second-Order Theories (SOTs) in which the non-ideal properties of the fluid are encoded in a new independent set of tensor variables whose evolution is determined by a Maxwell-Cattaneo-type relaxation dynamics. The corresponding relaxation time $\tau$ is a dimensionful parameter that ensures the finiteness of the perturbation propagation velocity and thus the hyperbolicity of the system \cite{romatschkeandromatschke,rezzollazanotti}. The typical Navier-Stokes behaviour is recovered after relaxation, or equivalently for $\tau\to0$.

In this work we choose SOTs to describe the fluid dynamics. In fact not only causality is ensured by the hyperbolic dynamics, but SOTs set a suitable framework for both the study of the evolution equations for general cases with arbitrary initial conditions and the numerical implementation due to the absence of spurious instabilities and the well posed initial value formulation \cite{romatschkeandromatschke,rezzollazanotti,hyl1983,hyl1985,rubio2018}. At the same time a common property of SOTs is the appearance of non-ideal spin-2 modes in the fluid description that are not present in FOTs. They can couple to GW to first order indeed capturing the viscous and free-streaming effects on the GW. Moreover, in \cite{herrerapavon2001,muronga2004} and references therein it was argued that SOTs are appropriate to study the relaxation processes of non-ideal fluctuations of the fluid into the equilibrium state, in our case, the ideal dominant radiation over time scales smaller than $\tau$.

Among all SOTs, we will work within the Divergence-Type Theory (DTT) framework \cite{geroch1,geroch2,reulanagy,estebanjeronimo1,estebanjeronimo2,estebanjeronimo3,nahuelaleesteban2020,lucas2020,calzetta2021} which has the advantage of fulfilling non-perturbatively the second law of thermodynamics. Nevertheless, since we assume first order perturbations around equilibrium we expect that the main effects qualitatively agree for most SOTs including the resumed BRSSS \cite{romatschkeandromatschke}, Anisotropic Hydrodynamics \cite{anhydro}, DNMR \cite{dnmr} and the Entropy Production Variational Principle \cite{estebanentropyprinciple}.

As we will see SOTs can be constructed from the Boltzmann equation with a particular parametrization of the one-particle distribution function (1pdf) \cite{lucas2020}. The new set of variables related to non-ideal properties of the fluid incorporates modes that are not present in the usual hydrodynamics called non-hydrodynamics modes \cite{romatschkeandromatschke}. Indeed it has been suggested that these modes might capture the nature of the relevant microscopic degrees of freedom for the non-ideal theory under consideration \cite{heller2015,romatschke2016,romatschkeandromatschke}.

We consider a specific cosmological scenario in which a standard inflationary de Sitter period brings every field into its vacuum state, except for the inflaton. Towards the end of Inflation, the inflaton reaches its minimum potential energy and begins to oscillate transfering its energy to all the matter fields in a process known as Reheating. In consequence the state of these matter fields becomes an excited many-particle state. In fact, quantum fluctuations for scales that were outside the horizon at the end of Inflation become larger than the local adiabatic vacuum ones and decohere. For these highly populated states a hydrodynamic description is suitable at large scales \cite{calzetta2008}.  For simplicity we assume an instantaneous Reheating \cite{peloso2006,lozanov2017,lozanov2019}. Finally a high-temperature thermal state composed by effectively massless quantum particles (the primordial plasma) is achieved at the beginning of the radiation dominated era. This state will be described by a real fluid with the background thermal radiation as a perfect fluid and small non-ideal (dissipative) fluctuations on top.

To give a concrete description of the primordial plasma, we consider that it is composed by all the Standard Model species plus an extra, weakly self-interacting, effectively massless and minimally coupled scalar field $s$. We will deal with very small masses and coupling constants and therefore this extra field would belong to the axion-like particles (ALP) family \cite{marsh2016}, frequently used in cosmology as dark matter or dark radiation candidates \cite{baumann2016,witten2017,bhupal2019,ghosh2020}. For simplicity we assume that, after Reheating, all the SM species form the main part of the background perfect fluid radiation. In turn the state of the scalar field $s$ will be effectively described as a viscous fluid in terms of its energy density (or temperature $T_s$), the four-velocity $u_\mu$ and a new tensor $\zeta_{\mu\nu}$ that takes into account the dissipative degrees of freedom. In other words we consider that the non-ideal fluctuations on top of the background radiation are entirely due to the hydrodynamic state composed by the particles of $s$. Since we are interested in analyzing the interaction between this fluid and the pGW we study linear perturbations around the background radiation. We implement a scalar-vector-tensor (SVT) decomposition of the degrees of freedom. We shall assume that the scalar and vector sector are thermalized with the rest of the dominant radiation state, and focus on the relaxation dynamics of the tensor spin-2 modes coming from $\zeta_{\mu\nu}$ coupled to the GW. This interaction defines the viscous effects of the primordial plasma on the pGW we are interested in.

Summarizing we assume a causal hydrodynamic description of the primordial plasma considering the non-ideal variables of the fluid as independent degrees of freedom as it is usual in SOTs. We define a concrete scenario in which the non-ideal properties are related to an extra effectively massless interacting scalar field. We analyze the propagation of the pGW through the viscous primordial plasma medium from the beginning of radiation era until today.

Since both the GW and the spin-2 modes of the fluid are independent variables, different sets of the initial condition can be considered and in particular we study two schemes. On the one hand we have the viscous effects channel  in which we use the usual initial spectrum for GW coming from quantum fluctuations during Inflation \cite{weinberg2008} and a vanishing one for the spin-2 modes of the fluid. This approach is related to the studies of the absorption of GW in presence of dissipative or free-streaming media in \cite{hawking1966,weinbergpaper1971,weinberglibro1972,anile1978,madore1973,esposito1971,prasanna1999} and more recently in \cite{weinberg2004,dent2013,baym2017,constraints_goswami2017,gravitonsandcdmarxivlu2018,brevik2019}. In particular we analyze the same conceptual issues discussed in \cite{weinberg2004,baym2017} where the GW-fluid dynamics was studied within a kinetic theory framework for both the collisionless (or free-streaming limit) and the collision-dominated regimes. Instead of the kinetic approach we use a self-consistent causal hydrodynamic scheme with independent non-ideal variables that captures the coupled dynamics between the GW and the fluid spin-2 modes to first order. Finally we explicitly compute the current spectrum of pGW.

On the other hand we have the GW production channel for which we assume vanishing initial GW and non-null initial fluid spin-2 modes. We extract these initial fluid spin-2 modes by matching the tensor part of the mean rms fluctuations of the energy momentum tensor at the beginning of radiation to the tensor part of the quantum noise kernel of the scalar field at the end of Inflation. In this way we analyze the production of GW due to the non-vanishing initial fluid spin-2 modes at the beginning of radiation and its later evolution until today. This evolution is characterized by the same dissipative dynamics considered before but applied to a non-vanishing initial condition for the fluid spin-2 fluctuations. It implies an effective decay of these fluctuations in a time scale $\tau$ coupled to GW. It is worth noting that fluctuations and dissipation naturally emerge from the evolution of interacting quantum fields \cite{esteban1995,esteban1997,calzetta2008}, therefore in our case we aim to capture the simplest dissipative dynamics within a hydrodynamic framework. Of course many other effects must be included in order to achieve a complete evolution. This part of the work could be related to studies about the production of GW by the thermalized plasmas \cite{delcampoford1988,ghiglieri2015,ghiglieri2020,mcdonough2020}. In the present case we do not explicitly consider thermal fluctuations \cite{nahuelaleesteban2020}, but we expect to include them in upcoming works.

The paper is organized as follows. In Section \ref{sec:primordialplasma} we elaborate on the primordial plasma description as a viscous fluid considering the thermal standard radiation as a perfect fluid background plus non-ideal perturbations. We define a causal hydrodynamical framework using a generalized DTT and we write down the dynamical equations for the mean energy density and for the spin-2 modes of the fluid. In Section \ref{sec:gwandtensors} we develop the coupled dynamics between the GW and the fluid spin-2 modes which takes into account the viscous effects. This evolution depends on background quantities that we introduce in Section \ref{settingthebackground}. The initial conditions and its numerical implementation are presented in Sections \ref{sec:initialspectra} and \ref{sec:gwandtensorsnumericalimplementation}. The resulting current spectrum of GW is described in Section \ref{sec:pgwspectrumtoday}. In particular we analyze the viscous effects channel in Section \ref{sec:pgwspectrumtodayviscouseffects} where we show the evolution of the pGW created during Inflation in presence of the viscous plasma, while the GW production channel due to the effective decay of the initial fluid spin-2 modes fluctuations is shown in Section \ref{sec:pgwspectrumtodayproductionofgwbythefluid}. Finally, in Section \ref{sec:conclusions}, we present the main conclusions of the work. We use the MTW convention for the metric signature $(-,+,+,+)$, $c=k_{\rm B}=\hbar=1$ and $M_{\rm pl}^2=1/(8\pi G)$.

\section{\label{sec:primordialplasma} Primordial plasma as a causal fluid}

We start by assuming that the primordial plasma is composed by all the relativistic degrees of freedom of the usual Standard Model fields plus an extra light (effectively massless) weakly-interacting minimally coupled scalar field, $s$, which represents a thermally produced ALP \cite{marsh2016,baumann2016,witten2017,bhupal2019,ghosh2020}. We describe the relativistic plasma as a background perfect fluid with a mean energy density $\rho_{\rm rad}$ and a rest frame four-velocity $U_{\rm rad}^\mu= \delta^\mu_0/a$, plus small fluctuations coming only from the new scalar field $s$. The mean energy density of the total radiation is
\bea
\rho_{\rm rad}=\rho_{\rm SM}+\rho_{s}=\frac{g_{*}}2 \,\rho_{\gamma}\label{energydensityplasma}
\tea
where $\rho_{\rm SM}$ and $\rho_s$ are the energy densities of the relativistic degrees of freedom of the Standard Model and the scalar field $s$ respectively. In addition $\rho_{\gamma}=\pi^2\,T_{\rm rad}^4/15$ is the mean energy density of photons with $T_{\rm rad}$ its physical temperature and $g_{*}(T_{\rm rad})$ the total number of relativistic degrees of freedom \cite{weinberg2008}. As we shall see the physical temperature of photons scales as $T_{\rm rad}\sim T_{\gamma,0} (a_0/a)$ due to the cuasi-adiabatic evolution of the Universe.

Our main assumption is that, after reheating, the extra light self-interacting scalar field $s$ becomes an effective viscous fluid in addition to the usual background ideal plasma of the Universe as we detail in Section \ref{sec:fluiddescription}. In this way we consider that all the non-ideal effects of the primordial plasma are due to this scalar field.

As we mentioned the appropriate theory which describes a non-perfect fluid in this cosmological context should be a covariant and causal hydrodynamic theory consistent with positive entropy production. We choose to work with SOTs since its dynamics is based on causal hyperbolic systems and it is appropriate to study the relaxation processes of non-ideal fluctuations of the fluid into the equilibrium state of ideal dominant radiation. We specifically use a DTT due to it ensures that the second law of thermodynamics is non-perturbatively fulfilled. Nonetheless we expect that most SOTs (not only DTTs) converge to the same qualitative behaviour to first order in perturbations. In reference \cite{lucas2020} a particular generalized DTT was developed and tested on Bjorken and Gubser flows with broad agreement with the well-known exact solutions coming from the kinetic theory. We take this generalized DTT to model our relativistic viscous fluid throughout the work.

It is useful to consider a conformal transformation in order to get rid of the expansion of the Universe. The conformal transformation is $g_{\mu\nu}=a^2\,\tilde g_{\mu\nu}=a^2\left(\eta_{\mu\nu}+h_{\mu\nu}\right)$ where the perturbation $h_{\mu\nu}$ represents the GW. It turns out that the DTT we use is conformally invariant as long as $p^\mu p_\mu =m_s^2\ll T_{{\rm phys},s}^2$ where $m_s$ is the mass of the scalar field $s$ and $T_{{\rm phys},s}$ the physical temperature of the fluid composed by $s$-particles. We assume $m_s$ is such that the massless regime holds throughout the evolution until today. Therefore we derive the comoving hydrodynamics equations for the effective fluid description of the scalar field $s$ in terms of its comoving temperature $T_s$, its four-velocity $u^\mu$ and the new tensor $\zeta^{\mu\nu}$ which captures the viscous effects.

\subsection{Microscopic description of the fluid}

We follow the prescription given in \cite{lucas2020} to extract the hydrodynamic equations which represent the effective fluid description of the self-interacting scalar field $s$. We start from kinetic theory by introducing the one-particle distribution function (1pdf) for massless scalar particles, namely the Bose-Einstein 1pdf. Since we want to describe non-ideal effects we include fluctuations on top of the equilibrium distribution function $f_0$. In turn, $f_0$ is written in terms of the mean temperature $T_s$ and the rest frame four-velocity of the fluid $U^\mu=\delta^\mu_0$ such that
\bea
f_0=d_s\;\frac{1}{\exp\left(-U^\mu p_\mu/ T_s\right)-1}\,,
\tea
where $d_s$ is the number of possible states of the scalar field $s$. We assume $d_s=2$ for convenience in view of future comparisons, nonetheless changing this parameter is straightforward. Fluctuations are introduced in the 1pdf through the temperature $\bm T_s=T_s+\delta T_s$, four-velocity $u^\mu=U^\mu+v^\mu$ and a non-equilibrium variable $\zeta_{\mu\nu}$ as
\bea
f=d_s\,\left[\exp\left(-u^\mu p_\mu/ \bm T_s-\frac{\zeta^{\mu\nu}p_\nu p_\mu}{\left(-\bm T_s u^\lambda p_\lambda\right)}\right)-1\right]^{-1}\label{1pdffull}
\tea
from which we extract the divergence (conservation) equations by taking moments of the Boltzmann equation.

We are interested in the dynamical evolution of the GW to first order in perturbations. In consequence we address a scalar, vector and tensor (SVT) decomposition of all degrees of freedom and we observe that the three sectors are decoupled from each other to first order. Thus we simplify our analysis by considering that the scalar and vector physical degrees of freedom, like the velocity and the temperature, are thermalized with the rest of the plasma at the very beginning of the radiation dominated era. We also assume that the scalar weak interactions between this $s$-fluid and the rest of the primordial plasma, that sustain the thermal equilibrium of the scalar sector, will eventually become inefficient due to the expansion of the universe at a temperature $T_{\rm rad}=T_{{\rm dec},s}$ and, in turn, the decoupling of the scalar degrees of freedom of the fluid occurs. The decoupling temperature $T_{{\rm dec},s}$ is an external parameter that we fix in Section \ref{sec:pgwspectrumtodayviscouseffects}. Indeed the ratio between the physical temperatures $T_{{\rm phys},s}/T_{\rm rad}=1$ until the decoupling of $s$. Afterwards $T_{{\rm phys},s}/T_{\rm rad}$ is determined by the conservation of entropy \cite{weinberg2008}. We elaborate on this later on when we compute the ratio between the energy density of $s$ and photons, $\rho_s/\rho_\gamma$. In consequence we only consider tensor (spin-2) fluctuations around the primordial plasma background state. At first order, they only can arise from the non-equilibrium variable $\zeta_{\mu\nu}$ of the fluid and the metric perturbation $h_{\mu\nu}$.

To include the viscous effects produced by the self-interaction between the spin-2 modes of the fluid we define an Anderson-Witting linear collision integral
\bea
I_{\rm col}=\frac{u^\mu p_\mu}{\tau} \left(f-f_0\right)\label{integralcollision}
\tea
where $\tau=\tau_{\rm phys}/a$ is the comoving relaxation time of the fluid for tensor spin-2 modes. In fact, from the quantum field theory perspective we would estimate the characteristic physical time of the self-interaction between spin-2 modes through a dimensionless coupling constant $g$ of the effectively massless scalar field as \cite{calzetta2008,berera1998}
\bea
\tau_{\rm phys}\sim\frac{1}{g^4\, T_{{\rm phys},s}}.\label{couplingconstanttau}
\tea
Considering that $T_{{\rm phys},s}/T_{\rm rad}\sim O(1)$, the estimation for the comoving relaxation time gives $\tau\sim 1/g^4\,T_{\gamma,0}$ with $T_{\gamma,0}$ the current photon temperature. The subscript $0$ means current values. For the collision-dominated regime it is possible to express an effective shear physical viscosity in terms of the relaxation time as $\eta_{\rm phys}\sim \rho_s \,\tau_{\rm phys}$ \cite{weinbergpaper1971,gavin1985,thoma1991}.

The idea of focusing all the non-ideal effects on just one extra field comes from the fact that only a dark matter candidate (or dark radiation for effectively massless fields) could have a value of $\tau$ large enough (or a small enough coupling constant) to obtain cosmologically relevant viscous effects. Among them, the ALPs turn to be a plausible choice \cite{marsh2016,baumann2016,witten2017,bhupal2019,ghosh2020}. Indeed we consider that $s$ is a thermally-produced weakly-interacting ALP \cite{baumann2016}, and in particular we are interested in very small coupling constant of the order $g\sim 10^{-4}-10^{-6}$ \cite{nahuelesteban2018} or very large relaxation times (analogous to the lifetimes) \cite{ghosh2020}. Assuming $T_{\gamma,0}\simeq 2.73\, {\rm K}$ and $a_0=1$, the comoving relaxation time reads
\bea
\tau\sim\frac1{g^4}\cdot 10^{-11}\,{\rm s}\,.
\tea

\subsection{\label{sec:fluiddescription}Causal hydrodynamics equations}

The hydrodynamic equations for a DTT consist in the vanishing divergence (conservation) of the current particle and the energy-momentum tensor both constructed through the usual procedure starting from kinetic theory \cite{lucas2020} (even used in extended geometries \cite{lescano2020}). In addition there exists an equation relating the divergence of the (new) non-equilibrium tensors to the collision integral (crf. Eq. (\ref{closure})) also derived from the Boltzmann equation.

These hydrodynamic equations are conformally invariant as long as $p^\mu p_\mu\simeq0$ (in fact $m_s\ll T_{{\rm phys},s}$) as shown in Appendix A of \cite{nahuelesteban2018}. Henceforth we use this invariance and we write down the equations in terms of the comoving variables with the metric $\tilde g_{\mu\nu}=\eta_{\mu\nu}+h_{\mu\nu}$. We only consider spin-2 perturbations to describe the first-order interaction between the spin-2 fluid modes coming from $\zeta_{\mu\nu}$ and the GW $h_{\mu\nu}$. The scalar and vector modes are not perturbed in the 1pdf (\ref{1pdffull}). Specifically the temperature and the four-velocity remain as zeroth-order variables determined by $T_s$ and $U^{\mu}={\delta^{\mu}}_0$ respectively. The first order quantities are $\zeta^{\mu\nu}$ and $h^{\mu\nu}$, which fulfill the following symmetry and gauge properties $\zeta_{\mu\nu}U^{\mu}={\zeta^{\mu}}_{\mu}=h_{\mu\nu}U^{\mu}={h^{\mu}}_{\mu}=0$.

As we have already mentioned we derive the dynamics following \cite{lucas2020} with the relevant integrals computed in \cite{aleesteban}. In this case the independent equations for the fluid are the conservation of the comoving energy momentum tensor
\bea
\tilde \nabla_\mu \tilde T^{\mu\nu}=0\label{consvervationtmunu}
\tea
with
\bea
\tilde T^{\mu\nu}=\int Dp\;p^\mu p^\nu\,f\label{tensort}
\tea
and the projected closure equation for the non-equilibrium tensors
\bea
{S^{\alpha\beta}}_{\mu\nu}\left[\tilde \nabla_\rho \tilde  A^{\mu\nu\rho}- \tilde K^{\mu\nu}-\tilde I^{\mu\nu}\right]=0\label{closure}
\tea
where
\bea
\tilde A^{\mu\nu\rho}=\int Dp\;p^\mu p^\nu p^\rho\,\left(\frac{1}{-U_\sigma p^\sigma}\right)\,f\,,\label{tensora}
\tea
\bea
\tilde K^{\mu\nu}=\int Dp\;p^\mu p^\nu \left[p^\lambda \tilde \nabla_\lambda\left(\frac{1}{-U_\sigma p^\sigma}\right)\right] f\label{tensork}
\tea
and
\bea
\tilde I^{\mu\nu}=\int Dp\;p^\mu p^\nu\left(\frac{1}{-U_\sigma p^\sigma}\right)\,I_{\rm col}\,.\label{tensori}
\tea

The quantity $p^\mu$ is the four-momentum and $Dp$ is the invariant momentum space integration measure defined as
\bea
Dp=\frac{2d^4p}{(2\pi)^3}\delta(p^2)\Theta(p^0)=\frac{d^3p}{(2\pi)^3p^0}\,.
\tea
The projector in equation (\ref{closure}) is
\bea
{S^{\alpha\beta}}_{\mu\nu}=\frac12\left[{\Delta^\alpha}_\mu\,{\Delta^\beta}_\nu+{\Delta^\alpha}_\nu\,{\Delta^\beta}_\mu-\frac23\Delta^{\alpha\beta}\,\Delta_{\mu\nu}\right]
\tea
with the spatial projector
\bea
\Delta^{\mu\nu}=\eta^{\mu\nu}+U^\mu U^\nu\,.
\tea
Recall that we are only considering tensor spin-2 perturbation to first order, thus the 1pdf $f$ in the expressions (\ref{tensort}), (\ref{tensora}) and (\ref{tensork}) is
\bea
f=f_0\Big[1+(1+f_0)\frac{\zeta^{\mu\nu}p_\mu p_\nu}{\left(-T_s\,U^\lambda p_\lambda\right)}\Big]
\tea
and the collision integral for equation (\ref{tensori}) is
\bea
I_{\rm col}=-f_0(1+f_0)\frac{\zeta^{\mu\nu}p_\mu p_\nu}{T_s\,\tau}\,.
\tea

The relevant quantities for studying the evolution of GW are the physical mean energy density of the fluid
\bea
\rho_s=\sigma\,T_{{\rm phys},s}^4,\label{densidadenergias}
\tea
with $\sigma=d_s\,\pi^2/30=\pi^2/15$, the transverse and traceless linearized comoving energy-momentum tensor in mixed components
\bea
{{\tilde T}^{(1)\,\mu}}{}_\nu{}^{\rm TT}=\frac8{15}\sigma \,T_s^4\,{\zeta^\mu}_{\nu}\label{t1munutt}
\tea 
and the linearized projected closure equation (\ref{closure})
\bea
{\zeta^{\mu\nu}}_{,0}+\frac1\tau \, \zeta^{\mu\nu}=-b \,{h^{\mu\nu}}_{,0}\,,\label{finalclosure}
\tea
the parameter $b$ depends on the specific DTT considered, $b=1/2$ in our case.

In summary the primordial plasma will be described by an energy momentum tensor with a perfect fluid background whose mean energy density is (\ref{energydensityplasma}) plus a viscous part which corresponds with the non-ideal linear fluctuations (\ref{t1munutt}) coming from the extra scalar field $s$. The dynamics of these spin-2 non-equilibrium linear fluctuations is given by (\ref{finalclosure}).

\section{Gravitational waves and fluid spin-2 modes dynamics}\label{sec:gwandtensors}

\subsection{Evolution equations}\label{sec:gwandtensorsevolutionequations}

The dynamical equations for the GW come from the tranverse (or divergenceless) and traceless (${\rm TT}$) projection of the linearized Einstein equations in mixed components, namely
\bea
{{G}^{(1)\,\mu}}{}_\nu{}^{\rm TT}=\frac{1}{a^2\,M_{\rm pl}^2}\,{{\tilde T}^{(1)\,\mu}}{}_\nu{}^{\rm TT}\label{einsteintensor}
\tea
where
${{G}^{(1)\,\mu}}{}_\nu{}^{\rm TT}$ and ${{\tilde T}^{(1)\,\mu}}{}_\nu{}^{\rm TT}$ the ${\rm TT}$ projections of the Einstein tensor and the comoving energy-momentum tensor of the primordial plasma respectively, both to first order. Due to the symmetries of $h_{\mu\nu}$ and $\zeta_{\mu\nu}$ we only consider the spatial components of (\ref{einsteintensor}) and then
\bea
{{G}^{(1)\,i}}{}_j{}^{\rm TT}=\frac12\left[-\eta^{\rho\sigma}\partial_\rho\partial_\sigma+2\frac{a'(\eta)}{a(\eta)}\partial_0\right]h_{ij},\label{tensoreinsteinorden1}
\tea
while ${{\tilde T}^{(1)\,\mu}}{}_\nu{}^{\rm TT}$ is given by the spatial components of (\ref{t1munutt}). Moreover, we expand the tensors in Fourier modes as
\bea
a^{\rm TT}_{ij}(\eta,\bm x)=\sum_{\lambda}\int \frac{d^3k}{(2\pi)^3}\,a_{\bm k,\lambda}(\eta)\,\epsilon^\lambda_{ij}(\bm k)\,e^{i\bm k \bm x}\label{TTmodesdecomposition}
\tea
in order to explicitly extract the spatial spin-2 (${\rm TT}$) modes. The component $a_{\bm k,\lambda}$ is the spin-2 Fourier mode with polarization index $\lambda$ corresponding to the polarization tensor $\epsilon^\lambda_{ij}(\bm k)$, for which $\epsilon_{ij}^\lambda(\bm k)k_i=\epsilon_{ii}^\lambda(\bm k)=0$ and $\epsilon_{ij}^\lambda(\bm k)\,\epsilon_{ij}^{\lambda'\,*}(\bm k)=2\delta_{\lambda \lambda'}$.

The GW dynamics is determined by the equation (\ref{einsteintensor}) and the expressions (\ref{tensoreinsteinorden1}) and (\ref{t1munutt}). Using the definition of the mean physical energy density of $s$ (\ref{densidadenergias}), of the critical energy density $\rho_c=3\,H^2 M_{\rm pl}^2\,$ and the Hubble constant $H=a'(\eta)/a^2(\eta)$, the TT projections of the Einstein equation (\ref{einsteintensor}) and of the closure equation (\ref{finalclosure}) turn out to be
\bea
h''_{{\bm k},\,\lambda}(\eta)+\frac{2a'(\eta)}{a(\eta)}h'_{{\bm k},\,\lambda}(\eta)+k^2h_{{\bm k},\,\lambda}(\eta)&=&6\,\left(\frac{\rho_s(\eta)}{\rho_{c}(\eta)}\right)\,\left(\frac{a'(\eta)}{a(\eta)}\right)^2\left[\frac{8\,\zeta_{{\bm k},\,\lambda}(\eta)}{15}\right].\label{hkeq}\\
\zeta'_{{\bm k},\,\lambda}(\eta)+\frac{1}{\tau}\,\zeta_{{\bm k},\,\lambda}(\eta)&=&-b\,h'_{{\bm k},\,\lambda}(\eta).\label{zkeq}
\tea
The scale factor $ a (\eta) $ and the densities $ \rho_c (\eta) $ and $ \rho_s(\eta) $ are fixed by the Friedmann equations and the conservation of entropy \cite{weinberg2008}. In this way it is possible to incorporate the effect of  $g_*(T_{\rm rad})$ into the dynamics obtaning a cuasi-adiabatic expansion of the Universe. As we will see including the decay of the relativistic degrees of freedom is relevant for the pGW evolution.

The system of equations (\ref{hkeq})-(\ref{zkeq}) represents the coupled linear dynamics between the spin-2 modes $\zeta_{{\bm k},\,\lambda}$ of the fluid which are related to the non-ideal effects and the pGW $h_{{\bm k},\,\lambda}$ (also spin-2 modes). The equation (\ref{hkeq}) is a well-known result in Cosmology and General Relativity which determines the production and evolution of the GW \cite{weinberg2008}. The right hand side acts as a source and it changes according to the different energy-momentum tensors considered. In the Introduction we mentioned several mechanisms could act as a source, if those phenomena were not correlated the total source would be the sum of each one.

It is remarkable that even in the case of vanishing source, the free propagation of the GW is not trivial due to the evolution of the scale factor $a(\eta)$. Several works elaborate on this point \cite{watanabe2006,kuroyanagi2009,saikawa2018} noting that the shape of the GW spectrum mainly depends on the background content of the Universe and, in particular, on the number of relativistic degrees of freedom $g_{*}$ related to the mean radiation energy density through the equation (\ref{energydensityplasma}). Naturally the scale factor $a(\eta)$ and the critical density $\rho_c(\eta)$ are determined by the Friedmann's equations, i.e. the zeroth order of the Einstein equations. In our case the source is the viscous part of the primordial plasma, specifically the spin-2 modes coming from (\ref{t1munutt}). As we can observe in (\ref{hkeq}) the ratio $\rho_s/\rho_c$, with $\rho_s$ the mean energy density of the scalar field state, measures the strength of the interaction between the spin-2 modes of the fluid and the pGW.

A robust and common behaviour of the linearized causal hydrodynamic theories, and in particular of the DTT considered in this work, is to provide a Maxwell-Cattaneo relaxation dynamics on the variables which describe the non-ideal effects over a time scale of order $\tau$ through an equation like (\ref{zkeq}). Interestingly enough, in Appendix \ref{kineticapproachappendix} we argue that this closure equation (\ref{zkeq}), typical of SOTs, recovers the same qualitative behaviour captured by the kinetic approach \cite{weinberg2004,baym2017} for both the collision-dominated and the collisionless regime. Instead, if we had used first order hydrodynamics to describe the non-ideal effects of the primordial plasma, we would have found that the right hand side of (\ref{hkeq}) would be proportional to $-\eta\, h_{\bm k,\lambda}'$ \cite{hawking1966,weinbergpaper1971,weinberglibro1972} and, of course, the equation (\ref{zkeq}) would not apply.

In a general Maxwell-Cattaneo relaxation dynamics it could be possible to define two independent time scales: $\tau_{\rm MC}$, related to the relaxation dynamics, and $\tau$, related to the viscous effects. Thus the closure equation reads
\bea
\zeta'_{{\bm k},\,\lambda}(\eta)+\frac1{\tau_{\rm MC}}\,\zeta_{{\bm k},\,\lambda}(\eta)&=&-b_{\rm eff}\,h'_{{\bm k},\,\lambda}(\eta)\label{zkeqgeneralmc2}
\tea
where $b_{\rm eff}=b\,\tau/\tau_{\rm MC}$ plays the role of an effective $b$-parameter. In the limit $\tau_{\rm MC}\to0$ we recover the first order hydrodynamics behaviour. Our case corresponds to $\tau_{\rm MC}= \tau$.

\subsection{Setting the background}\label{settingthebackground}

To solve the system (\ref{hkeq})-(\ref{zkeq}) we first find $a(\eta)$ by numerically integrating the Friedmann's equation
\bea
H^2=\frac{1}{3\,M_{\rm pl}^2}\Big(\rho_{\rm rad}+\rho_{\rm M}+\rho_{\Lambda}\Big)
\tea
in a fiducial spatially flat cosmological background determined by the current density parameters and the current Hubble constant. The values of the cosmological parameters are close to those obtained by Planck \cite{planck2018}. We assume $H_0=h\,100$ km/s/Mpc with $h=0.7$ and the density parameters for dust matter and dark energy are $\Omega_{\rm M}=0.3$ and $\Omega_\Lambda=1-\Omega_{\rm rad}-\Omega_{\rm M}\simeq 0.7$. For radiation we have the equation (\ref{energydensityplasma}). On the one hand we define the density parameter of the photonic radiation today which is $\Omega_{\gamma}=5.04\cdot10^{-5}$ and on the other hand we have $g_*$ that takes into account the photonic and non-photonic radiation. In particular we need the evolution of the relativistic degrees of freedom as a function of the plasma temperature
\bea
g_{*}(T_{\rm rad})=g_{*{\rm SM}}(T_{\rm rad})+g_{{*}s}(T_{\rm rad})
\tea
where $g_{*{\rm SM}}$ and $g_{{*}s}$ correspond to the Standard Model fields and the extra scalar field $s$. We extract the detailed evolution of $g_{\rm *SM}(T_{\rm rad})$ from reference \cite{saikawa2018}. On the other hand, since $s$ is effectively massless until today its physical mean energy density, considering two degenerated states, reads $\rho_s=\pi^2\,T_{{\rm phys},s}^4/15$. Therefore $g_{*s}=2\,\rho_s/\rho_{\gamma}$ is determined by the ratio of the physical temperatures $T_{{\rm phys},s}/T_{\rm rad}$. At the very beginning of the radiation dominated era, where the temperatures of all fields are in equilibrium, $T_{{\rm phys},s}=T_{\rm rad}$. After decoupling at temperatures $T_{\rm rad}\leq T_{{\rm dec},s}$,
\bea
\frac{T_{{\rm phys},s}}{T_{\rm rad}}=\left[\frac{g_{*,{\rm ent}}(T_{\rm rad})}{g_{*,{\rm ent}}(T_{{\rm dec},s})}\right]^{1/3}\label{tstrad}
\tea
due to the entropy conservation as usual. In the expression (\ref{tstrad}) $g_{*\,{\rm ent}}(T_{\rm rad})$ means the relativistic degrees of freedom related to the entropy density, not to the energy density. We also read this quantity from \cite{saikawa2018}.

In addition we obtain the cuasi-conformal relation between $T_{\rm rad}$ and $a$ from the entropy conservation due to the cuasi-adiabatic expansion of the Universe, i.e. $g_{*\,{\rm ent}}(T_{\rm rad})\,a^3\,T_{\rm rad}^3={\rm const.}$ \cite{weinberg2008}. We find the initial state at the beginning of radiation dominated era by the backward evolution from today until a reheating temperature $T_\gamma\simeq 6\cdot10^{15}$ GeV. Since we assume Reheating occurs in no time and with no loss of energy, it implies that $H_{\rm inf}\simeq5\cdot10^{13}$ GeV giving a tensor-to-scalar parameter $r\simeq0.04$ \cite{weinberg2008}.

Since the spin-2 modes of the fluid are independent variables we must proceed to set the initial conditions for both the pGW and these spin-2 modes in the following section.

\subsection{Initial spectra}\label{sec:initialspectra}

Since we assume an instantaneous Reheating we set the initial conditions of the evolution at a time $\eta=\eta_I$ by matching quantum fluctuations at the end of Inflation with the stochastic fluctuations at the onset of the radiation dominance.

\subsubsection{Gravitational waves}

During Inflation we consider the usual framework for the GW \cite{weinberg2008,reviewcaprinifigueroa2018} where the two independent polarization amplitudes $h_{\bm k,\lambda}$, with $\lambda=+,\times$, are regarded as two canonical massless scalar quantum fields $\tilde h_{\bm k,\lambda}=M_{\rm pl}\,a(\eta)\,h_{\bm k,\lambda}/\sqrt{2}$. We assume the Bunch-Davies vacuum state for these fields and therefore, after canonical quantization, we are able to compute $\langle 0| \tilde h_{\bm k,\lambda}\,\tilde h_{\bm k',\lambda'}|0\rangle$. Next, we use the Landau prescription to obtain the stochastic expectation values from the quantum ones. This implies that $\langle A B\rangle_{\rm S}=1/2 \langle \{A; B\}\rangle_{\rm Q}$ where $\{\cdot\,;\cdot\}$ is the anticommutator. We get the pGW spectrum for super-horizon scales ($k<a_IH_I$)  
\bea
\left\langle h_{\bm k,\lambda}\,h^*_{\bm k',\lambda'} \right\rangle_{\rm S}=\frac12\,\left\langle 0|\left\{ h_{\bm k,\lambda}\,;\,h^*_{\bm k',\lambda'}\right\}|0\right\rangle=(2\pi)^3\,\delta_{\lambda \lambda'}\,\delta(\bm k-\bm k')\,\left(\frac{2\pi^2}{2k^3}\right)\,\mathcal P_h(k)\label{hquantumvev},
\tea
with
\bea
\mathcal P_h(k)=\frac{H_{\rm inf}^2}{\pi^2\,M_{\rm pl}^2}\Bigg|_{k=aH},\label{spectrumpgravitons}
\tea
evaluated at the time where the mode with wavenumber $k$ exits the horizon. We are mostly interested in the evolution of the stochastic background of GW, so we take the initial spectra in the simplest inflationary scenario where $H_{\rm inf}={\rm const.}$ and in consequence ${\mathcal P}_h(k)$ is scale invariant.

\subsubsection{Fluid spin-2 modes}

In this section we set the initial conditions for the non-equilibrium tensor modes $\zeta_{\bm k,\lambda}$ by relating the effective hydrodynamic fluctuations of the primordial plasma at the beginning of radiation dominance to the vacuum quantum fluctuations at the end of Inflation. For this purpose we will match the self-correlation of the energy-momentum (noise kernel) on both sides of the transition. During the radiation era we have the stochastic fluctuations of the fluid and during Inflation we have the quantum fluctuations of the scalar field $s$ in its de Sitter vacuum \cite{nahuelesteban2018}. See Appendix \ref{appendixtensorfluid} for details.

On the one hand we consider the spatial TT projection of the noise kernel for the scalar field $s$ during Inflation. It reads
\bea
\left[N^{i\,\,\,k}_{\,\,\,j\,\,\,l}(x, x')\right]^{\rm TT}=\frac12\, \Big\langle 0\Big|\left\{ \left(\hat {T^{i}}_j(x)-\left\langle \hat {T^{i}}_j(x)\right\rangle\right);\left(\hat {T^{k}}_l(x')-\left\langle \hat {T^{k}}_l(x')\right\rangle\right)\right\} \Big|0\Big\rangle^{\rm TT},\label{noisekernel1}
\tea
where $|0\rangle$ is the Bunch-Davies vacuum and $\hat {T^i}_j$ are the spatial components of the energy momentum of the scalar field $s$. The noise kernel for a minimally coupled effectively massless scalar field in de Sitter space was computed in \cite{guillem2010,guillemtesis,desittercorrelator2020}.

We are interested in the classicalized large scales that are outside the horizon at the end of Inflation ($k<a_IH_{\rm inf}$). In consequence we take a renormalized (or classicalized) noise kernel by subtracting the local adiabatic vacuum for scales inside the horizon ($k>a_I H_{\rm inf}$) as it has been done in \cite{nahuelesteban2018}. As a result we obtain a vanishing noise kernel for the scales inside the horizon while keeping unchanged the behaviour for outer scales. Unlike reference \cite{nahuelesteban2018}, here the horizon plays the explicit role of a physical ultraviolet cutoff. In particular the classicalized noise kernel (\ref{noisekernel1}) at $\eta=\eta_I^-$ becomes
\bea
\left[N^{i\,\,\,k}_{\,\,\,j\,\,\,l}(\bm x,\bm x', \eta_I)\right]_{\rm Q}^{\rm TT}\simeq\int \frac{d^3k}{(2\pi)^3}\,e^{i\bm k(\bm x-\bm x')}\,\Lambda^{ijkl}(\bm k)\frac{H_{\rm inf}^5}{5\,\pi^3\,a_I^3}\;\Theta\left(a_IH_{\rm inf}-k\right)\label{noisekernelq}\,,
\tea
where $a_I=a(\eta_I)$, $\Lambda^{ijkl}(\bm k)$ is the spatial TT projector and Q means quantum expectation value at the end of Inflation.

On the other hand we have the TT projection of the energy-momentum tensor self-correlation of the effective fluid to first order in perturbations which reads
\bea
\left[N^{i\,\,\,k}_{\,\,\,j\,\,\,l}(\bm x,\bm x', \eta_I)\right]_{\rm S}^{\rm TT}=\Big[\left\langle {T^{(1)\;i}}_j(\bm x){T^{(1)\;k}}_{l}(\bm x')\right\rangle-\left\langle {T^{(1)\;i}}_j(\bm x)\right\rangle\left\langle{T^{(1)\;k}}_{l}(\bm x')\right\rangle\Big]_{\eta=\eta_I}^{\rm TT},
\tea
where ${T^{(1)\;k}}_{l}$ is the physical version (not comoving) of the energy-momentum tensor to first order (\ref{t1munutt}) and S means stochastic average at the onset of radiation dominated era. Using $\left\langle{T^{(1)\;k}}_{l}\right\rangle=0$ we get
\bea
\left[N^{i\,\,\,k}_{\,\,\,j\,\,\,l}(\bm x,\bm x', \eta_I)\right]_{\rm S}^{\rm TT}&=&2\left(\frac{8\sigma}{15}\right)^2\,T_\gamma^8\,\left\langle {\zeta^{{\rm TT}\;i}}_j(\bm x){\zeta^{{\rm TT}\;k}}_{l}(\bm x')\right\rangle_{\eta=\eta_I}=\nn
&=&\int \frac{d^3k}{(2\pi)^3}e^{i\bm k(\bm x-\bm x')}\Lambda^{ijkl}(\bm k)\,2\left(\frac{8\sigma}{15}\right)^2\left(\frac{2\pi^2}{2k^3}\right)\,T_\gamma^8\; {\cal P}_{\zeta}(k,\eta_I)\label{noisekernels}\,,
\tea
here $T_\gamma$ is the reheating physical temperature determined by $H_{\rm inf}^2=g_{*}(T_\gamma)T_\gamma^4\,\pi^2/(90M_{\rm pl}^2)$.

By matching (\ref{noisekernelq}) and (\ref{noisekernels}), we find that
\bea
{\mathcal P}_{\zeta}(k)=\frac{1}{23040\,\pi}\frac{\left.g^2_{*}\right|_{\eta_I}}{\sigma^2}\left(\frac{H_{\rm inf}}{M_{\rm pl}}\right)^4\left(\frac{k}{a_I\,H_{\rm inf}}\right)^3\;\Theta\left(a_IH_{\rm inf}-k\right).\label{spectrumpfluid}
\tea
Note that this spectrum reaches its highest values for scales near the cutoff $k\lesssim a_IH_{\rm inf}$ but it is globally suppressed by the scale-independent factor $(H_{\rm inf }/M_{\rm pl})^4$.

Summarizing, the initial spectrum ${\mathcal P}_\zeta(k,\eta_I)$ for $\zeta_{\bm k,\lambda}$ represents the hydrodynamic spin-2 fluctuations of the effective fluid describing the interacting excited state of $s$ at the beginning of radiation dominated era. We extract the expression ${\mathcal P}_\zeta(k,\eta_I)$ from the non-vanishing quantum noise kernel of $s$ at the end of Inflation. Here we explicitly relate the spectrum of the non-equilibrium spin-2 modes $\langle\zeta^2_{\bm k,\lambda}\rangle$ and the TT projection of the self-correlation of the scalar field energy-momentum tensor. In consequence, this framework allows us to study the evolution of these noise kernel fluctuations of the scalar field $s$ after Inflation, when an effective hydrodynamic state is achieved.

\subsection{Numerical implementation}\label{sec:gwandtensorsnumericalimplementation}

We rewrite the equations (\ref{hkeq})-(\ref{zkeq}) with the change $u\rightarrow k\eta$ and we get
\bea
h_{\bm k}''(u)+\frac{2\,a'(u)}{a(u)}\,h_{\bm k}'(u)+h_{\bm k}(u)&=&6\left(\frac{\rho_s(u)}{\rho_{c}(u)}\right)\left(\frac{a'(u)}{a(u)}\right)^2\left[\frac{8\,\zeta_{\bm k}(u)}{15}\right]\label{eqforh}\\
\zeta_{\bm k}'(u)+\frac{1}{k\,\tau}\,\zeta_{\bm k}(u)&=&-b\,h_{\bm k}'(u)\label{eqforzeta},
\tea
for each polarization mode. Hereafter a prime ($'$) denotes a derivative with respect to $u$. 

The stochastic backgrounds of the primordial GW and the spin-2 modes of the fluid were originated from quantum fluctuations during Inflation, which are classicalized after the corresponding modes exit the horizon. Therefore our analysis will be valid for scales that are outside the horizon at the end of Inflation ($k<a_IH_{\rm inf}$), eventually these scales re-enter the horizon at late times.

Our next goal is to implement the simplest initial conditions that apply cosmologically \cite{weinberg2004} in the linear system of equations (\ref{eqforh})-(\ref{eqforzeta}). Therefore we propose the following ansatz
\bea
h_{\bm k}(u)=h_{\bm k}^{\rm prim}\,h_1(k,u)+\zeta_{\bm k}^{\rm prim}\,h_2(k,u)\label{hmode}\\
\zeta_{\bm k}(u)=h_{\bm k}^{\rm prim}\,z_1(k,u)+\zeta_{\bm k}^{\rm prim}\,z_2(k,u)
\tea
which distinguishes the stochastic variables $h^{\rm prim}_{{\bm k},\,\lambda}$ and $\zeta^{\rm prim}_{{\bm k},\,\lambda}$, regarding the initial primordial spectra, from the transfer functions $h_i( k,\eta)$ and $z_i(k,\eta)$, regarding the dynamical evolution. It implies that the transfer functions are constrained by \cite{weinberg2004}
\bea
&&h_1(k,u_I)=1\;\;\;\;\;\;h'_1(k,u_I)=0\;\;\;\;\;\;z_1(k,u_I)=0\label{ic1}\\
&&h_2(k,u_I)=0\;\;\;\;\;\;h'_2(k,u_I)=0\;\;\;\;\;\;z_2(k,u_I)=1\label{ic2}.
\tea
There exists a third combinations of the initial conditions that corresponds to an independent solution which rapidly decays and becomes negligible with respect to the previous ones and in consequence we do not consider it. The primordial spectra fulfill
\bea
\left\langle h_{{\bm k},\lambda}^{\rm prim}\,h^{{\rm prim}*}_{{\bm k'},\lambda'}\right\rangle&=&(2\pi)^3\delta_{\lambda \lambda'}\,\delta(\bm k-\bm k')\frac{2\pi^2}{2k^3}\,\mathcal P_h(k,\eta_I)\label{statprophh}\\
\left\langle \zeta_{{\bm k},\lambda}^{\rm prim}\,\zeta^{{\rm prim}*}_{{\bm k'},\lambda'}\right\rangle&=&(2\pi)^3\delta_{\lambda \lambda'}\,\delta(\bm k-\bm k')\frac{2\pi^2}{2k^3}\,\mathcal P_\zeta(k,\eta_I)\label{statpropzz}\\
\left\langle h_{{\bm k},\lambda}^{\rm prim}\,\zeta^{{\rm prim}*}_{{\bm k'},\lambda'}\right\rangle&=&\left\langle \zeta_{{\bm k},\lambda}^{\rm prim}\,h^{{\rm prim}*}_{{\bm k'},\lambda'}\right\rangle=0\label{statprophz},
\tea
with $\mathcal P_h(k,\eta_I)$ and $\mathcal P_\zeta(k,\eta_I)$ given in (\ref{spectrumpgravitons}) and (\ref{spectrumpfluid}) respectively. The initial cross correlation is vanishing because both spectra correspond to different physical phenomena and indeed they are uncorrelated. Replacing the ansatz in (\ref{eqforh})-(\ref{eqforzeta}) and using the statistical properties of the primordial spectra (\ref{statprophh})-(\ref{statprophz}) we obtain that the coupled equations for the transfer functions $h_{1,2}$ and $z_{1,2}$ are
\bea
h_i''(k,u)+\frac{2\,a'(u)}{a(u)}\,h_i'(k,u)+h_i(k,u)&=&\frac{16}5\left(\frac{\rho_s(u)}{\rho_{c}(u)}\right)\left(\frac{a'(u)}{a(u)}\right)^2\,z_i(k,u)\label{transferh}\\
z_i'(k,u)+\frac{1}{k\,\tau}\,z_i(k,u)&=&-b\,h_i'(k,u)\,.\label{transferz}\tea
Using the initial conditions (\ref{ic1}) we obtain the solutions for $h_1$ and $\zeta_1$, instead the initial conditions (\ref{ic2}) determine $h_2$ and $z_2$.

We numerically solve the equations (\ref{transferh})-(\ref{transferz}) for both initial conditions (\ref{ic1}) and (\ref{ic2}). We use a method based on \cite{saikawa2018}. The quantities $a$, $\rho_s$ and $\rho_c$ were described in Section \ref{settingthebackground}.

\section{Current pGW spectrum}\label{sec:pgwspectrumtoday}

In this section we describe the current GW spectrum. The GW energy density is
\bea
\rho_{\rm GW}(t)&=&\frac{M_{\rm pl}^2}{4}\,\left\langle \dot h_{ij}(t,\bm x)\,\dot h^{ij}(t,\bm x)\right\rangle\nonumber\\
&=&\frac{M_{\rm pl}^2}{4}\,\frac{1}{a^2(\eta)}\,\int d\left(\log k\right)\,\sum_{\lambda}\,\left[\mathcal P_h(k)\left[h'_1(k,\eta)\right]^2+\mathcal P_\zeta(k)\left[h'_2(k,\eta)\right]^2\right],
\tea
where we use the decomposition (\ref{TTmodesdecomposition}) and the expression (\ref{hmode}). Nonetheless the observationally relevant quantity is the current ($\eta=\eta_0$) spectrum of the GW density parameter defined by
\bea
\Omega_{\rm GW}(k,\eta_0)&=&\frac{1}{\rho_c}\frac{d\rho_{\rm GW}}{d\log k}=\Omega_{{\rm GW},h}(k,\eta_0)+\Omega_{{\rm GW},\zeta}(k,\eta_0)\label{omega}
\tea
with
\bea
\Omega_{{\rm GW},\,h}(k,\eta_0)&=&\frac{1}{6\,a_0^2\,H_0^2}\,\mathcal P_h(k)\left[h'_1(k,\eta_0)\right]^2\,,\label{omegah}\\
\Omega_{{\rm GW},\,\zeta}(k,\eta_0)&=&\frac{1}{6\,a_0^2\,H_0^2}\,\mathcal P_\zeta(k)\left[h'_2(k,\eta_0)\right]^2\,.\label{omegazeta}
\tea

The spectrum $\Omega_{\rm GW}(k,\eta)$ has two uncorrelated contributions coming from the stochastic primordial spectra ${\mathcal P}_h$ and ${\mathcal P}_\zeta$. The transfer functions $h_1$ and $h_2$ have the same dynamical equations but with different initial conditions (eqs. (\ref{ic1})-(\ref{ic2})).

\subsection{Viscous effects channel}\label{sec:pgwspectrumtodayviscouseffects}

In this section we focus our attention on the contribution $\Omega_{{\rm GW},\,h}(k,\eta_0)$  (eq. (\ref{omegah})) for which the initial spectrum of the GW is ${\mathcal P}_h$ and the initial spin-2 modes of the fluid are vanishing.

To describe a representative example we have to fix the coupling constant $g$, which in turn sets the relaxation scale $\tau$ through (\ref{couplingconstanttau}), and the decoupling temperature $T_{{\rm dec},s}$. It turns out that $g$ and $\tau$ define  the dissipation scale. We take $g\simeq 10^{-6}$ because the associated frequency $f_\tau=1/2\pi \tau\sim 10^{-14}$ Hz (and wavelength $\lambda_\tau\sim 1 $ Mpc) could be of cosmological interest. Note that the wavenumber $k$, wavelength $\lambda$ and frequency $f$ are comoving quantities, and since we choose the convention $a_0=1$ they correspond to their physical values today. Further they are related through $f=ck/2\pi=c/\lambda$ (hereafter we explicitly reintroduce the speed of light, $c$, for clarity). At the same time the decoupling temperature determines the function $\rho_s/\rho_c$ and the effective number of non-photonic radiation, $N_{\rm eff}$. The latter quantity is a decreasing function of the decoupling temperature. In consequence we choose the minimum value, $T_{{\rm dec},s}=125$ MeV, for which $N_{\rm eff}\simeq 3.5$ (or $\Delta N_{\rm eff}\lesssim0.5$) saturates the current 3$\sigma$-bound (99.7 \% CL) from both the CMB \cite{planck2018} and the BBN \cite{olive2020,ghosh2020} separately. Since in a recent work \cite{olive2020} several statistical combinations of both measurements gave tighter constraints, we explain how our results change for lower values of $N_{\rm eff}$ at the end of this section. There we also mention the impact of the variation in $\tau$ (or $g$). Finally as the thermally produced scalar field particles are effectively massless until today we do not expect an appreciable impact on the evolution of the structure formation process during the matter-dominated era.

\begin{figure}
	\centering
	\includegraphics[width=15cm]
	{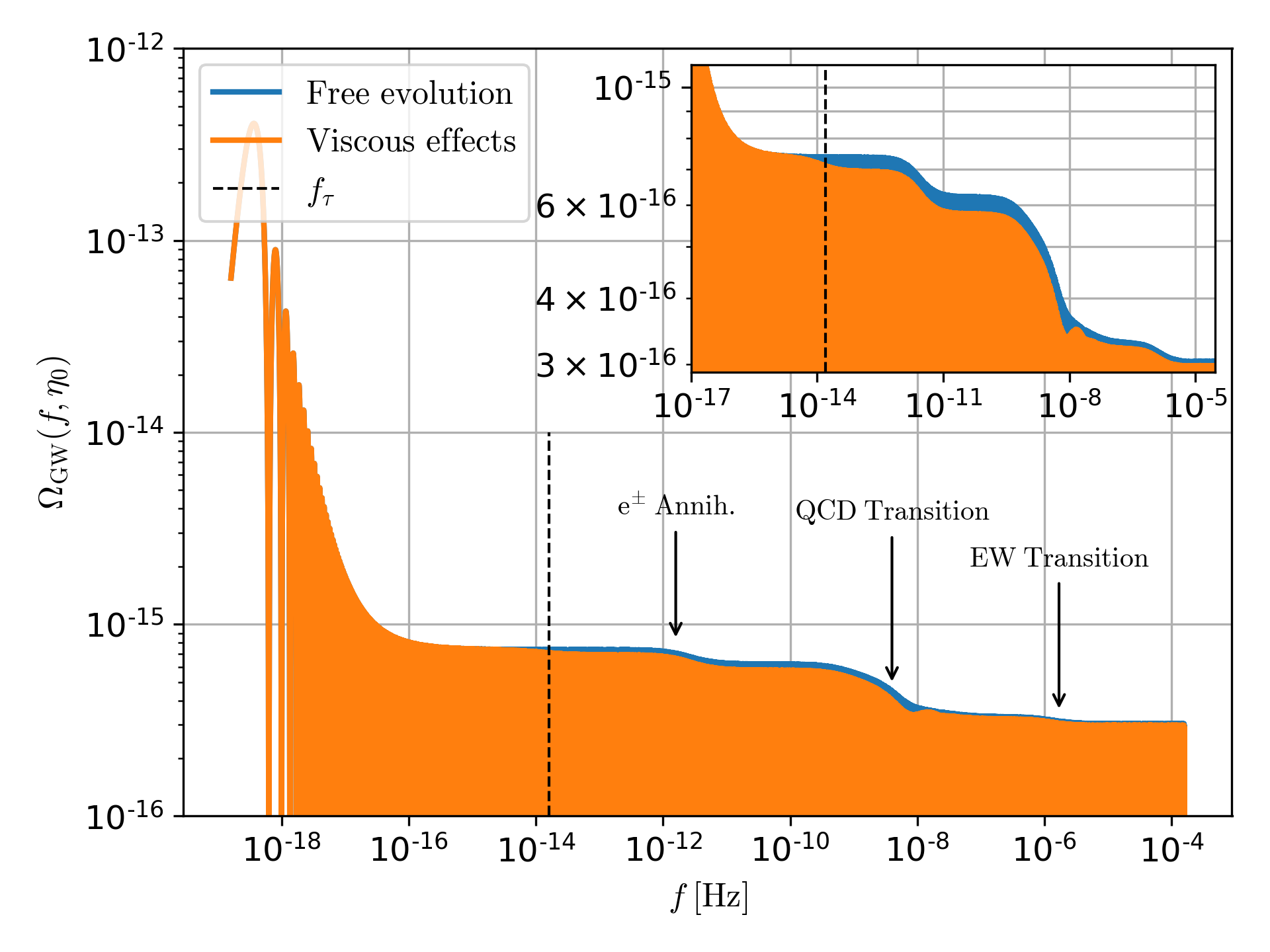}
	\caption{Current pGW spectrum $\Omega_{{\rm GW},h}(f,\eta_0)$ (eq. (\ref{omegah})). The blue solid line represents the free evolution (i.e. with no source) and the orange one takes into account the viscous effects of the primordial plasma on the pGW evolution. We observe a limiting frequency $f_\tau=1/2\pi\tau$ related to the characteristic time scale $\tau$ of the dissipative phenomena. For the collision-dominated regime ($f/f_\tau<1$) the impact of the viscous effects erases the fluid anisotropies and the source effects on the GW is negligible. Instead for the collisionless regime ($f/f_\tau>1$) the GW-fluid interaction produces a damping of the GW amplitude driven by the expansion of the Universe, the loss of the GW energy leads to a relative decrease of the spectrum amplitude of about 10 \%. The smooth steps on both spectra (blue and orange) are related to the decay of the different relativistic degrees of freedom.}
	\label{fig:spectra100}
\end{figure}

In Fig. \ref{fig:spectra100} we show the results of the numerical integration for $\Omega_{{\rm GW},\,h}(k,\eta_0)$. First we present the current spectrum of pGW for free evolution, i.e. when no source is considered (blue solid line in Fig. \ref{fig:spectra100}). In that case we recover the well-known spectrum of pGW \cite{watanabe2006, saikawa2018, kuroyanagi2009}, actually with slight differences due to the incorporation of the extra scalar field $s$. It is characterized by two regimes: one for scales that reenter the horizon after equality in the matter dominated era, $k<k_{\rm EQ}$ and another for scales reentering the horizon before equality while the universe is radiation dominated, $k>k_{\rm EQ}$, with the scale $k_{\rm EQ}\sim 10^{-2}/{\rm Mpc}$ related to the horizon size at equality. For $k<k_{\rm EQ}$ the amplitude scales as $k^{-2}$ and for $k>k_{\rm EQ}$ we find an almost flat spectrum. In fact, the smooth steps are related to the decay of the different relativistic degrees of freedom as it is shown in Fig. \ref{fig:spectra100}. The change to dark energy domination is only observable for very large wavelength $k<k_{\Lambda}\sim 10^{-5}/{\rm Mpc}$ and it is not relevant for this analysis.

When the interaction with the fluid is present (orange solid line in Fig. \ref{fig:spectra100}) we observe that the limiting frequency $f_\tau$ distinguishes two regimes: the collision-dominated regime, $f/f_\tau \ll 1\to kc\tau\ll1$, where interactions become efficient and the collisionless regime $f/f_\tau \gg 1\to kc\tau\gg1$, characterized by the free-streaming fluid spin-2 modes. It turns out that for the collision-dominated stage, the interactions tends to erase the fluid anisotropies and in turn the GW source becomes negligible \cite{baym2017} while in the collisionless or free-streaming stage the anisotropic tensor is non-negligible and in consequence the damping of GW driven by the expansion of the Universe arises \cite{weinberg2004,baym2017}. In Appendix \ref{kineticapproachappendix} we elaborate on both regimes and we present a comparison between our SOT scheme and the kinetic approach showing a broad qualitative agreement.

\begin{figure}
	\centering
	\includegraphics[width=15cm]
	{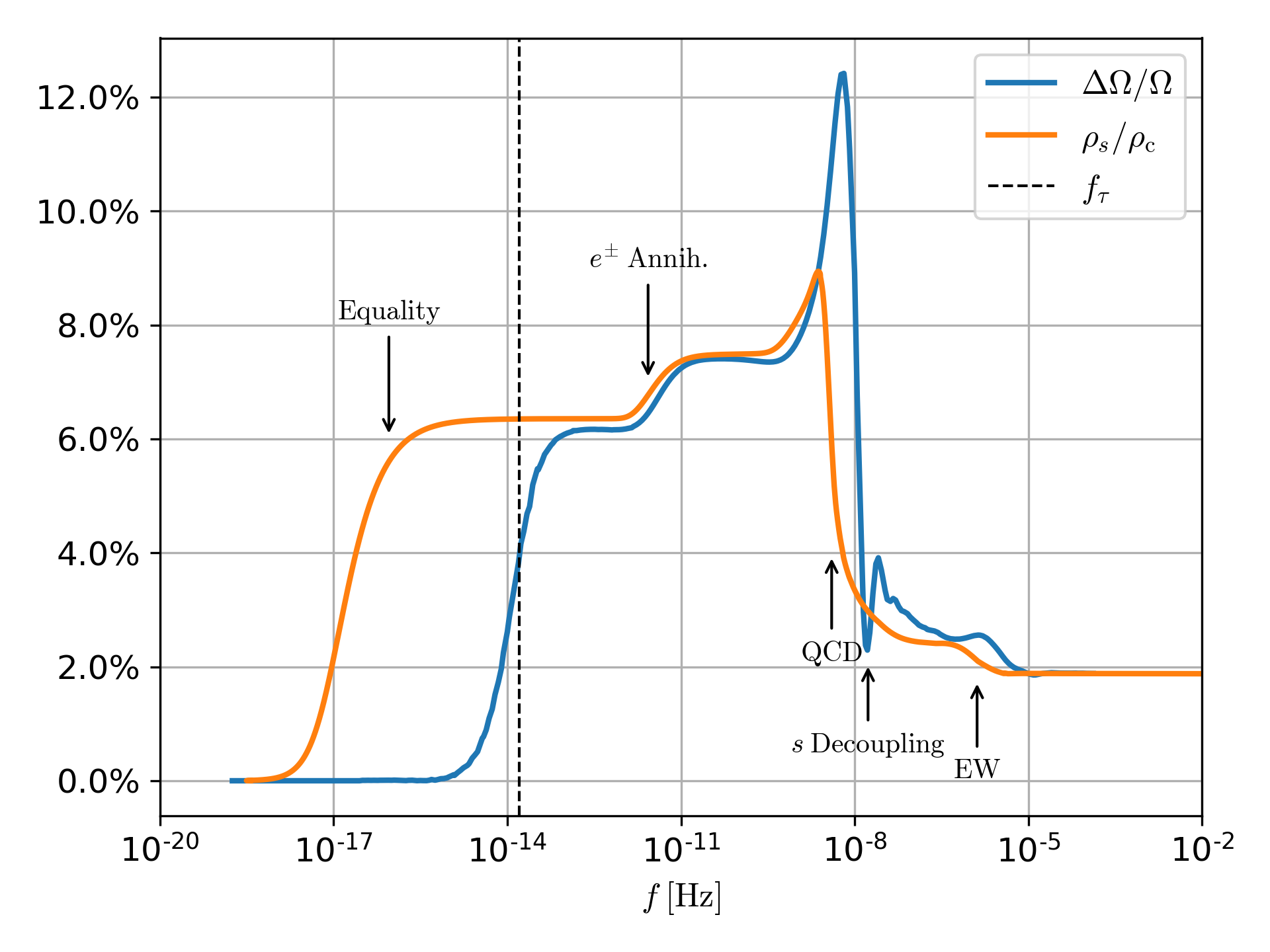}
	\caption{Relative difference $\Delta \Omega/\Omega$ (eq. (\ref{deltaomega})) between the two pGW spectra presented in Fig. \ref{fig:spectra100}. It measures the impact of the viscous effects with respect to the free evolution case. In addition we show the ratio $\rho_s/\rho_c$ between the mean energy density of the scalar fluid and the critical density of the Universe, as a function of the frequency corresponding to the scale that equals the horizon size at different times through the relation $f=H(\eta)a(\eta)/2\pi$. We observe that the relative supression of the spectrum of pGW in the collisionless regime ($f>f_\tau$) is of about 10 \% and it is strongly correlated with the ratio $\rho_s/\rho_c$. However it is negligible for long wavelengths ($f<f\tau$) independently of $\rho_s/\rho_c$. The wiggly feature around $s$ decoupling is a numerical artifact that comes from the assumption of a sudden decoupling. In this case we consider $N_{\rm eff}\simeq3.5$ ($\Delta N_{\rm eff}\lesssim0.5$) at the edge of the observational 3$\sigma$-constraints from the CMB \cite{planck2018} and the BNN \cite{olive2020,ghosh2020} separately. For values in $3.1<N_{\rm eff}<3.5$ we find the same qualitative behaviour with maximum relative suppression that gradually runs from 2\% to 12\%.}
	\label{fig:deltaomega}
\end{figure}

The viscous and the damping effects are mainly determined by the scale $k_\tau=1/c\tau$ and the ratio $\rho_s/\rho_c$ which defines the strength of the GW-fluid interaction. It is possible to estimate an effective physical viscosity related to this fluid for the collision-dominated regime as $\eta_{{\rm phys}}=(8b/15)\,\rho_{s}\,\tau_{{\rm phys}}$. Its current value $\eta_{{\rm phys},0}$ is negligible compared with the bounds corresponding to the interstellar medium viscosity that might affect the propagation of GW produced e.g. during a black-holes collision \cite{constraints_goswami2017,gravitonsandcdmarxivlu2018}. To quantify the effects of the non-ideal processes with respect to the free evolution spectrum we define the relative difference between both, namely
\bea
\frac{\Delta \Omega}{\Omega}(k)=\frac{\left|\Omega_{\rm GW, free}(k)-\Omega_{\rm GW, non-ideal}(k)\right|}{\Omega_{\rm GW, mean}(k)}.\label{deltaomega}
\tea
In Fig. \ref{fig:deltaomega} we show the quantity $\Delta \Omega/\Omega$ vs. $f$ ($f=ck/2\pi$). For the collisionless regime ($f>f_\tau$) the relative difference is of about 10\%. As we have argued, this depletion in the amplitude is strongly correlated to the ratio $\rho_s/\rho_c$, however for the collision-dominated regime ($f<f_\tau$) the interactions becomes efficient, isotropizes the fluid and wipe the GW source out regardless of $\rho_s/\rho_c$ converging to the free evolution spectrum. To construct the curve $\rho_s/\rho_c$ vs. $f$ we relate the conformal cosmic time $\eta$ with the particular scale which equals the horizon size at that moment through the condition $k=H(\eta)a(\eta)$.

Beyond this example we analyze a wide range of parameters. For all characteristic frequencies $f_\tau$, fixed by any coupling constants within our values of interest $g\sim10^{-6}-10^{-4}$, we find the existence of the two above-mentioned regimes: the collision-dominated regime, $f<f_\tau$, with no absorption of GW and the collisionless one, $f>f_\tau$, with a characteristic GW damping effect due to the free-streaming spin-2 modes of the fluid. On the other hand, decoupling temperatures $T_{{\rm dec},s}\geq 125\,{\rm MeV}$ imply that $3.1\leq N_{\rm eff}\leq 3.5$. The minimum value $N_{\rm eff}=3.1$  is reached for any temperature $T_{{\rm dec},s}>m_{\rm top}\sim200\,{\rm GeV}$. The qualitative effect and the distinction of regimes, reported in Fig. \ref{fig:spectra100} and \ref{fig:deltaomega}, remain the same for the entire range of $N_{\rm eff}$ but with a maximum value of $\Delta\Omega/\Omega$ gradually running from 12\% to 2\% when $N_{\rm eff}$ goes from 3.5 to 3.1.

Finally we conclude that the contribution $\Omega_{{\rm GW},\,h}(k,\eta_0)$ to the total spectrum (\ref{omega}) represents the non-trivial evolution of the pGW, created by quantum fluctuations during Inflation, in the presence of the viscous primordial plasma. It is related to both purely dissipative and collisionless effects, particularly the loss of the GW energy through the interaction with the fluid spin-2 modes occurs in the collisionless limit.

\subsection{GW production channel}\label{sec:pgwspectrumtodayproductionofgwbythefluid}

\begin{figure}
	\centering
	\includegraphics[width=15cm]
	{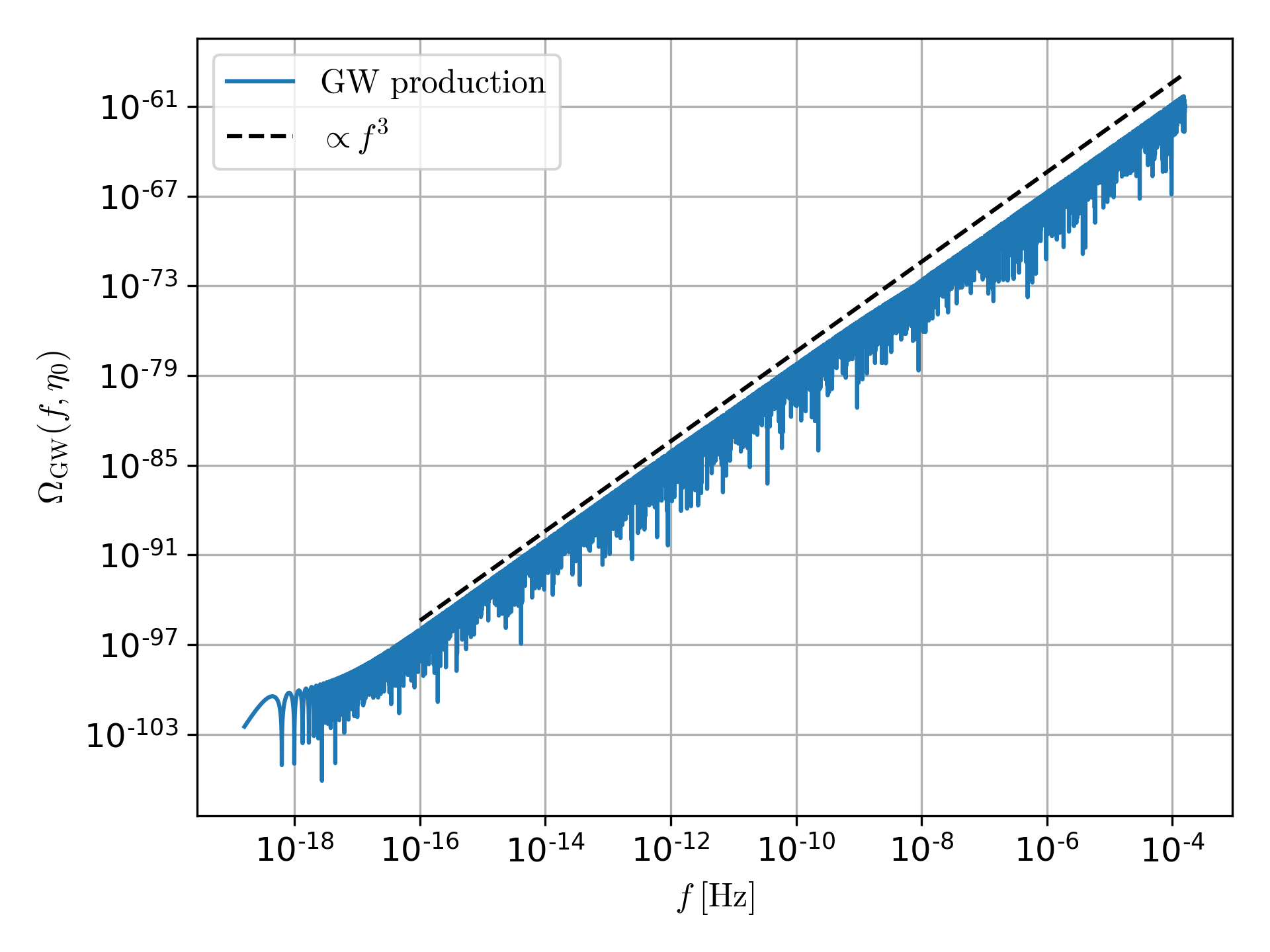}
	\caption{Current spectrum of GW entirely produced by the effective decay of the initial spin-2 modes of the fluid, $\Omega_{{\rm GW},\zeta}(k,\eta_0)$ (eq. (\ref{omegazeta})). This initial fluctuations are related to the noise kernel of the scalar field at the end of Inflation. The amplitude of the spectrum is globally suppressed by a factor $(H_{\rm inf}/M_{\rm pl})^4$. Since the transfer function turns out to be cuasi-flat for small wavelengths, the spectrum scaling mainly depends on the initial spectrum $\mathcal{P}_\zeta(k)\sim k^3$. For the range of frequencies considered this spectrum is negligible compared with the one of Fig. \ref{fig:spectra100}. Nonetheless, this spectrum might be relevant for very high frequencies ($f\sim10^8$ Hz) corresponding to the horizon size at the end of Reheating.}
	\label{fig:spectra001}
\end{figure}

In this section we analyze the contribution $\Omega_{{\rm GW},\,\zeta}(k,\eta_0)$ (eq. (\ref{omegazeta})). We show the resulting spectrum in Fig. \ref{fig:spectra001}. Since we start the evolution with vanishing initial GW, this spectrum is entirely produced by the initial spin-2 modes of the fluid, which are related to the (TT projection) noise kernel of the scalar field $s$ at the end of Inflation. The transfer function $z_2(k,\eta)$ (eqs. (\ref{ic2}) and (\ref{transferz})) represents the effective decay dynamics of these noise kernel fluctuations coupled to the GW after Inflation where the scalar field many-particle state is effectively described as a fluid.

The amplitude of the spectrum $\Omega_{{\rm GW},\,\zeta}$ is strongly suppressed by a factor $(H_{\rm inf}/M_{\rm pl})^4$. Even though the transfer function $h_2(k,\eta)$ turns out to be non-trivial, it is cuasi-flat for small wavelengths so the scaling of $\Omega_{{\rm GW},\,\zeta}$ mainly depends on the initial spectrum $\mathcal{P}_\zeta(k)\sim k^3$. Therefore the current total production of GW by the fluid, $\Omega_{{\rm GW},\zeta}(k,\eta_0)$, becomes negligible compared with $\Omega_{{\rm GW},\,h}(k,\eta_0)$ for the frequency range we are considering.

In consequence we conclude that for frequencies of cosmological interest, the total spectrum of pGW $\Omega_{\rm GW}(k,\eta_0)$ (eq. (\ref{omega})) is basically determined by $\Omega_{{\rm GW},\,h}(k,\eta_0)$ which is shown in Fig. \ref{fig:spectra100}.

The spectrum $\Omega_{{\rm GW},\,\zeta}(k,\eta_0)$ could be relevant for very high frequencies related to the horizon size at the end of Reheating, $f\sim 10^8$ Hz. It worth noting that the final result depends on the regularization and renormalization method used to compute ${\mathcal P}_\zeta$ (eq. (\ref{spectrumpfluid})) which is sensitive to the energy scale of Inflation $H_{\rm inf}$.

\section{Conclusions}\label{sec:conclusions}

We have described the linear evolution of the pGW including the viscous effects of the primordial plasma within a covariant and causal hydrodynamic framework (based on a generalized DTT) for which the entropy production fulfills the second law of thermodynamics non-perturbatively.

We consider a standard cosmological scenario with an extra light self-interacting scalar field, belonging to the ALP family, that becomes a many-particle state at the onset of radiation and it is effectively described as a causal real (viscous) fluid. In this generalized DTT-framework the set of hydrodynamic degrees of freedom is extended in order to capture dissipative phenomena. The dynamics is described by hyperbolic equations which ensure causality. It also allows to take into account relaxation phenomena on time scales of order $\tau$ or smaller. Among this extended set of variables there are spin-2 modes related to the non-ideal effects which are coupled to the pGW (also spin-2) to first order. In consequence we focus our attention on these spin-2 modes of the fluid, $\zeta$, and we write down the linear equations that determine the coupled dynamics between the pGW and $\zeta$, namely equations (\ref{eqforh})-(\ref{eqforzeta}).

We find the current spectrum of the GW density parameter $\Omega_{\rm GW}(k,\eta_0)$ (eq. (\ref{omega})) by numerically solving the system (\ref{eqforh})-(\ref{eqforzeta}) with the initial conditions (\ref{hmode})-(\ref{statprophz}). This spectrum has two uncorrelated contributions.

On the one hand we have $\Omega_{{\rm GW},h}(k,\eta_0)$ corresponding to the viscous effects channel that represents the evolution of the usual pGW, created by quantum fluctuations during Inflation, in presence of a relativistic viscous plasma. The viscous effects mainly come from the part of the fluid composed by the effectively massless self-interacting scalar field particles and are effectively characterized by a dimensionful parameter $\tau$ through a linear collision integral (relaxation approximation eq. (\ref{integralcollision})). From Fig. \ref{fig:spectra100} we observe a limiting frequency $f_\tau=1/2\pi\tau$ related to the viscous effects that distinguishes two regimes the collision-dominated, $f/f_\tau<1$, and the collisionless one, $f/f_\tau>1$. In the former the collisions are efficient and isotropize the fluid erasing its ability to source GW, in consequence we obtain negligible effects on the pGW spectrum. In the latter an absorption of the GW energy occurs due to the damping effect produced by the free-streaming fluid spin-2 modes and driven by the expansion of the Universe. 

It turns out that by virtue of the typical closure equation of SOTs (\ref{zkeq}), which couples to the GW thorugh equation (\ref{hkeq}) and includes its back-reaction on the fluid, we recover the same qualitative behaviour reported in works where the GW-fluid interaction is described using a kinetic theory approach \cite{weinberg2004,baym2017} (see Appendix \ref{kineticapproach}). In this work we set a simple and concrete realization of the fluid viscous effects and explicitly compute the current spectrum of pGW. The parameters were fixed in order to analyze scales of cosmological interest. In particular the coupling constant $g$ between the fluid spin-2 modes determines the values of $\tau$ and the characteristic frequency $f_\tau$. For very small coupling constant $g\simeq10^{-6}$, this frequency is $f_\tau\simeq 10^{-13}$ Hz. The amplitude decrease with respect to the free-evolution spectrum for small wavelengths depends on the number of non-photonic radiation $N_{\rm eff}$ determined basically by the decoupling temperature $T_{{\rm dec},s}$. The maximum relative decrease runs from 2 to 12 \% according $N_{\rm eff}$ ranges from 3.1 to 3.5.

On the other hand $\Omega_{{\rm GW},\zeta}(k,\eta_0)$ is related to the production of GW by the decay of the initial spin-2 modes of the fluid coming from the TT projected noise kernel fluctuations of the scalar field $s$ at the end of Inflation. This contribution is negligible compared with $\Omega_{{\rm GW},h}(k,\eta_0)$ for the scale considered and it could only be relevant for very high frequencies ($10^8$ Hz) related to the end of Reheating. This simple approach allows studying the production of GW given by the effective decay of the spin-2 fluctuations of the fluid. Thus, for example, it would be interesting to include thermal fluctuations in this formalism \cite{nahuelaleesteban2020} and to analyze the GW production \cite{delcampoford1988,ghiglieri2015,ghiglieri2020,mcdonough2020}. We expect to address this topic in future works. Further studies could consider other initial spectra ${\mathcal P}_\zeta$ owing to different physical phenomena in order to look for an enhancement of the GW spectrum at different scales.

Finally it would also be interesting to apply this causal viscous hydrodynamic scheme to study the effect of dissipation on the production of GW in at least two scenarios: in a binary neutron-star merger \cite{rezzollaprl2017}, and in a cosmological phase transition due to the sound waves or turbulent motion of the fluid resulting from the expansion or the collisions of bubbles. For the latter it is particularly relevant to model the dissipative-like effective interaction between the relativistic plasma and the field which develops the symmetry breaking phenomenon \cite{hindmarsh2017,leitao2016}.

\begin{acknowledgments}
	I would like to thank Esteban Calzetta for many enlightening discussions. I am deeply indebted for his encouragement, support and advice. In the same way, conversations about this manuscript with Alejandra Kandus, Eric Lescano and Andrés Perez were very helpful and pleasant. I also thank to Departamento de Física, Facultad de Ciencias Exactas y Naturales, Universidad de Buenos Aires for financial support. Work supported in part by Universidad de Buenos Aires through grant 20020170100129BA.
\end{acknowledgments}

\appendix

\section{Stochastic noise kernel during Inflation}\label{appendixtensorfluid}

Let us consider an exact de Sitter Inflation and a minimally coupled scalar field in its Bunch-Davies vacuum state. The noise kernel reads
\bea
N_{\mu\nu\rho'\sigma'}(x,x')=\frac12\,\Big\langle 0\Big|\Big\{ \big(T_{\mu\nu}(x)-\big\langle  T_{\mu\nu}(x)\big\rangle\big);\big( T_{\rho'\sigma'}(x')-\big\langle T_{\rho'\sigma'}(x')\big\rangle\big)\Big\}\Big|0\Big\rangle\,,\label{noisekernel_appendix}
\tea
where $|0\rangle$ is the Bunch-Davies vacuum state and $T_{\mu\nu}(x)$ is the energy-momentum tensor of the scalar field. The noise kernel (\ref{noisekernel_appendix}) in de Sitter space were computed in terms of the Wightman function $G_{xx'}=\left\langle0\right|\hat\phi(x)\hat\phi(x')\left|0\right\rangle$ in references \cite{guillem2010,guillemtesis,desittercorrelator2020,huverdaguer2008}. Several issues about the well-defined expression of the quantum energy-momentum tensor have to be considered, e.g. the point-splitting regularizaton. It turns out that the noise kernel is a well-defined quantity with an expected divergence for the coincidence limit ($x\to x'$). Since we are interested in the spatial tensor part of the noise kernel, we only consider the contribution of the kinetic term to the energy-momentum tensor. Although the effectively massless limit is not equivalent to the exact massless case, the TT projection avoids this singular behaviour. In turn, the spatial components of the noise kernel coming from the kinetic term read
\bea
N^{i\,\,\,\,k'}_{\,\,\,\,j\,\,\,\,l'}\,(x,x')&=&\frac12\Big[g^{in}g^{k'm'}\nabla_n\nabla_{m'}G_{xx'}\nabla_j\nabla_{l'}G_{xx'}+g^{in}g^{l'm'}\nabla_n\nabla_{m'}G_{xx'}\nabla_j\nabla_{k'}G_{xx'}+\nn
&&+(x\leftrightarrow x')\Big]\label{noisekernelspatialappendix}\,.
\tea
In \cite{synge1960} there are explicit formulae to compute derivatives of bitensors (tensors evaluated at two different point).

As we have mentioned in Section \ref{sec:initialspectra} we will consider that the noise kernel fluctuations classicalize (or freeze out) upon horizon exit as usual during Inflation. In fact, it is possible to achieve a \textit{renormalized} noise kernel by subtracting the local adiabatic vacuum fluctuations for the inside-horizon scales \cite{nahuelesteban2018}. Finally the \textit{renormalized} noise kernel for equal times at the end of Inflation $\eta=\eta'=\eta_I$ reads
\bea
N^{i\,\,\,\,k}_{\,\,\,\,j\,\,\,\,l}\,(r)&=&\frac{H_{\rm inf}^4}{8\,\pi^4\,a_I^4}
\Big[r^ir^jr^kr^lF_1(r)+(\delta^{il}r^jr^k+\delta^{jk}r^ir^l)F_2(r)+\nn
&&+\delta^{il}\delta^{jk}F_3(r)+(k\leftrightarrow l)\Big]\label{noisekerneltouse}
\tea
where
\begin{equation}
F_1(r)=\frac{4}{r^8}\,\Theta\Big(r-r_0\Big)\label{f1r}
\end{equation}
\begin{equation}
F_2(r)=-\frac{2}{r^6}\,\Theta\Big(r-r_0\Big)\label{f2r}
\end{equation}
\begin{equation}
F_3(r)=\frac{1}{r^4}\,\Theta\Big(r-r_0\Big)\label{f3r}
\end{equation}
with $r^i=(\bm x^i-\bm x'^{\,i})$ means the components of the comoving spatial coordinates and $r_0=\alpha/a_IH_{\rm inf}$ with $\alpha\sim1$. In the right hand side of the expression (\ref{noisekerneltouse}) we drop the prime in the indexes and we lower and raise indexes in the spatial sector with the Kronecker-delta.

The TT projection of (\ref{noisekerneltouse}) is
\bea
\left[N^{i\,\,\,\,k}_{\,\,\,\,j\,\,\,\,l}\,(r)\right]^{\rm TT}&=&\int \frac{d^3p}{(2\pi)^3}\,e^{i\bm p(\bm x- \bm x')}\,\frac{H_{\rm inf}^4}{4\,\pi^4\,a_I^4}\,\Lambda^{ijkl}(\bm p)\,\times\nonumber\\
&&\times\left[\frac{2\hat F_1^{\prime\prime}(p)}{p^2}-\frac{2\hat F_1^{\prime}(p)}{p^3}-\frac{2\hat F_2^{\prime}(p)}{p}+ \hat F_3(p)\right]\label{ttprojectionnoisekernel}
\tea
with
\bea
\hat F_i(p)=\int d^3r\, e^{-i\bm p \bm r}F_i(r)\,,
\tea
and the transverse and TT projectors in Fourier space
\bea
P^{ij}(\bm k)&=&\delta^{ij}-\frac{k^ik^j}{k^2}\\
\Lambda^{ijkl}(\bm k)&=&\frac12\left[P^{ik}(\bm k)P^{jl}(\bm k)+P^{il}(\bm k)P^{jk}(\bm k)-P^{ij}(\bm k)P^{kl}(\bm k)\right]\,.
\tea

After computing the Fourier transforms of (\ref{f1r})-(\ref{f3r}) and replace them on (\ref{ttprojectionnoisekernel}) we obtain the expression (\ref{noisekernelq}).

\section{Collision-dominated and collisionless regimes: comparison with the kinetic approach}\label{kineticapproachappendix}

In reference \cite{baym2017} the authors generalize the work of \cite{weinberg2004} in order to introduce the effect of collisions in the evolution of GW coupled to a fluid, both using a kinetic theory framework. Let us analyze the compatibility of this kinetic theory framework with our treatment using a SOT for the hydrodynamic description of the fluid.

\subsection{Minkowski background}

Let us first analyze the evolution of GW coupled to a causal fluid in a Minkowski background with the metric being
\bea
ds^2=-dt^2+\left(\delta_{ij}+h_{ij}\right)dx^i dx^j\,,
\tea
where $h_{ij}$ represents the GW and $h^{i}{}_i=\partial_i h_{ij}=0$. The dynamic equations turns out to be
\bea
\left(\frac{\partial^2}{\partial\,t^2}-\nabla^2\right) h_{ij}(\bm r,t)=\frac{2}{M_{\rm pl}^2}\,T^{(1)}_{ij}{}^{\rm TT}(\bm r,t)\,.\label{ecuacionh}
\tea

We use the generalized DTT described in the main text in order to model the fluid and therefore the matter source of GW, i.e. the spin-2 part of the EMT to first order, is defined as
\bea
T^{(1)}_{ij}{}^{\rm TT}=\frac{8}{15}\,\rho_s\,\zeta_{ij}\label{ecuacionzeta}\,,
\tea
with $\rho_s$ the mean energy density of the fluid. The dynamics of the new independent non-equilibrium variable $\zeta_{ij}$ is fixed by the equations (\ref{closure}) and (\ref{finalclosure}) in which we assume the relaxation time approximation for the collision integral. Indeed we obtain
\bea
\frac{\partial}{\partial t}\zeta_{ij}+\frac{1}{\tau}\zeta_{ij}=-b\frac{\partial}{\partial t}h_{ij}\label{cierre}\,.
\tea
Applying the Fourier transform to the equations (\ref{ecuacionh}) and (\ref{cierre}) in both space and time and using the decomposition (\ref{TTmodesdecomposition}) we get, for each polarization,
\bea
\left(-\omega^2+k^2\right)h_{\bm k,\,\omega}&=&\frac{16}{15}\frac{1}{M_{\rm pl}^2}\,\rho_s\,\zeta_{{\bm k},\,\omega}\,\\
\left(i\omega-\frac1{\tau}\right)\zeta_{{\bm k},\,\omega}&=&-b\,i\,\omega h_{\bm k,\,\omega}\,.
\tea

The general dispersion relation is
\bea
\omega^2-k^2-\frac{B}{1+i/(\omega\tau)}=0
\tea
with $B=16\,\rho_s\,b/(15\,M_{\rm pl}^2)$. In the collisionless or free-streaming limit,  $\omega\tau\gg1$, the dispersion relation reads
\bea
\omega^2-k^2-B\left(1-\frac{i}{\omega\tau}\right)=O\left[\left(\frac{1}{\omega\tau}\right)^2\right]\,.\label{disprelcollisionless}
\tea
For the collision-dominated regime where collisions are efficient, $\omega\tau\ll1$, the dispersion relation is
\bea
\omega^2-k^2+i\,B\,\omega\tau=O\left[\left(\omega\tau\right)^2\right]\label{disprelcollisiondom}\,.
\tea
Both expressions (\ref{disprelcollisionless}) and (\ref{disprelcollisiondom}) qualitatively agree with the results presented in \cite{baym2017}, in particular we observe that the damping exponent increases linearly with $\omega\tau$ in the collision-dominated regime while it decreases as $(\omega\tau)^{-1}$ for the collisionless limit giving the maximum damping for $\omega \tau\sim1$. It is important to note that more collisions in a fluid imply a lower value of $\tau$.

\subsection{Cosmological background}

The case described in this work is related to the evolution of the GW coupled to matter in a cosmological background instead of static Minkowski one. This has already been studied in \cite{weinberg2004} for the free-streaming limit and generalized in \cite{baym2017} in order to incorporate the effects of collisions, both within a kinetic theory framework. In this appendix we qualitatively compare their main results with our SOT hydrodynamic approach.

For the simplest SOT, the main dynamical equations are (\ref{eqforh}) and (\ref{eqforzeta}). Let us analyze the viscous effects channel (Section \ref{sec:pgwspectrumtodayviscouseffects}) in which we assume that the initial spin-2 modes of the fluid are vanishing, i.e. $\zeta_{\bm k}(u_I)=0$. This is equivalent to set an equilibrium initial 1pdf in presence of GW as in \cite{weinberg2004,baym2017}. In fact it is possible to cast the system of equations (\ref{eqforh})-(\ref{eqforzeta}) as the following simple formal equation
\bea
h_{\bm k}''(u)+\frac{2\,a'(u)}{a(u)}\,h_{\bm k}'(u)+h_{\bm k}(u)&=&-24\,\frac{\rho_s(u)}{\rho_{c}(u)}\left(\frac{a'(u)}{a(u)}\right)^2\int_{u_I}^ud\tilde u\,e^{-(u-\tilde u)/(k\tau)}\,\frac{h'_{\bm k}(\tilde u)}{15}\quad\label{eqforhtotalapendice}
\tea
where the right side is the total matter source of GW with no initial spin-2 modes of the fluid. The exponential factor takes into account collisions and two limiting regimes can be analyzed. In the collision-dominated regime, $ck\tau\ll1$, we observe that collisions are efficient and erase the fluid anisotropies turning the right hand side of (\ref{eqforhtotalapendice}) to be negligible due to the exponential factor $e^{-(u-\tilde u)/(k\tau)}$, independently of $\rho_s/\rho_c$ and $(a'/a)^2$.  For the collisionless limit $ck\tau\gg1$ we obtain that the matter source of GW yields an amplitude decrease in the GW spectrum produced by a generalized damping driven by the expansion of the universe (see Fig. \ref{fig:spectra100}). In this way we recover the qualitative behaviour reported in \cite{weinberg2004,baym2017}.

In the kinetic approach, considering collisions, the right hand side of (\ref{eqforhtotalapendice}) reads \cite{baym2017}
\bea
-24\,\frac{\rho_s(u)}{\rho_{c}(u)}\left(\frac{a'(u)}{a(u)}\right)^2\int_{u_I}^ud\tilde u\,e^{-(u-\tilde u)/(k\tau)}\,K(u-\tilde u)\,h'_{\bm k}(\tilde u)\label{kineticapproach}
\tea
with $K(s)=j_2(s)/s^2$ and $j_2(s)$ the spherical Bessel function of second order. The kernel $K(s)$ corresponds to the effect of spatial fluid anisotropies produced by particles moving along the propagation direction of the GW. 

Our equation (\ref{eqforzeta}) for the spin-2 modes of the fluid is analogous to the linearized equation for the perturbation of the 1pdf in the kinetic approach. It leads to the equation (\ref{eqforhtotalapendice}) that effectively captures the same qualitative dynamics of the kinetic approach as we mentioned in the previous paragraphs. The momentum average performed to construct the equation (\ref{eqforzeta}) to first order (see equations (\ref{closure})-(\ref{tensori})) and the chosen parametrization of the 1pdf do not allow us to express the effect of the kernel $K(s)$ and in consequence we obtain a trivial one $K(s)\to K(0)=1/15$ as we observe by comparing the right hand side of (\ref{eqforhtotalapendice}) and (\ref{kineticapproach}).


\bibliographystyle{myJHEP}


\bibliography{references_pgw_2020}

\end{document}